\documentclass[fleqn,12pt,twoside]{article}

\usepackage{graphicx}
\usepackage[figuresright]{rotating}

\newcommand{\AmS}{{\protect\the\textfont2
  A\kern-.1667em\lower.5ex\hbox{M}\kern-.125emS}}

\title{The Nucleon Elastic Form Factors}

\author{Donal Day\address{Department of Physics
   University of Virginia, Charlottesville, Virginia 22904 }}
       
\usepackage{graphicx}
\usepackage{amsmath}
\usepackage{espcrc1}
\usepackage{pifont}
\usepackage{times}

\newcommand{\gen}{$G_E^n\,$}
\newcommand{\gmn}{$G_M^n\,$}

\begin{document}

\maketitle

\begin{abstract}
   The nucleon elastic form factors are of fundamental interest, provide a unique testing ground for QCD motivated models of nucleon structure and are of critical importance to our understanding of the electromagnetic properties of nuclei.
   Even after an experimental effort spanning nearly 50 years the nucleon form factors are still the subject of active investigation.
   Advances in polarized beams, polarized targets and recoil polarimetry have been exploited over the last decade to produce an important and precise set of data.
I  review the status of the experimental efforts to measure the nucleon elastic form factors.
\end{abstract}

\section{INTRODUCTION}

The nucleon elastic form factors encapsulate important information about  their internal structure and have been the subject of sustained experimental and theoretical investigations for almost 50 years. Recent experimental efforts, especially those exploiting spin degrees of freedom, have produced an impressive data set out to large momentum transfers, generating considerable theoretical interest.



The earliest investigations exploited elastic electron scattering to determine the Dirac and Pauli form factors, $F_1$ and $F_2$, which are functions of momentum transfer, $Q^2=4E_0E'\sin^2(\theta/2)$, alone. In single photon exchange, the elastic cross section is written as:
\begin{equation}
\frac{d \sigma}{d \Omega}   =  \sigma _{\rm{Mott}}\frac{E'}{E_{0}}\left\{ \left( F_{1}\right) ^{2}+\tau \left[ 2\left( F_{1}+F_{2}\right) ^{2}\tan ^{2}\left( \theta\right) +(F_{2}^{})^{2}\right] \right\}, \label{eq:f1f2}
 \end{equation}
 where  $\tau=\frac{Q^2}{4M^2}$, $\theta$ is the electron scattering angle, $E_0,E'$ are the incident and final electron energies respectively, and $\sigma _{\rm{Mott}}$ is the Mott cross section. Because of their  direct relation (in the Breit frame) to the Fourier transforms of the charge and magnetization distributions in the nucleon, the Sachs form factors are commonly used. They are  linear combinations of $F_1$ and $F_2$: 
$G_{E}   =  F_{1}-\tau F_{2}$ and 
$G_{M}   =  F_{1} +F_{2}$. Early measurements of the form factors established a scaling law relating three of the four nucleon elastic form factors and the dipole law describing their common $Q^2$-dependence, $G_{E}^{p}(Q^2)  \approx \frac{G_{M}^{p}(Q^2)}{\mu_{p}}\approx \frac{G_{M}^{n}(Q^2)}{\mu_{n}} \approx  G_{D}\equiv \left( 1+Q^{2}/0.71\right) ^{-2}.$ This behavior is known as form factor scaling.

\section{EARLY EXPERIMENTAL TECHNIQUES}
\subsection{Proton Form Factors}
\begin{figure}
  \begin{center}
    \begin{minipage}[c]{0.48\textwidth}
      \includegraphics[width=\textwidth]{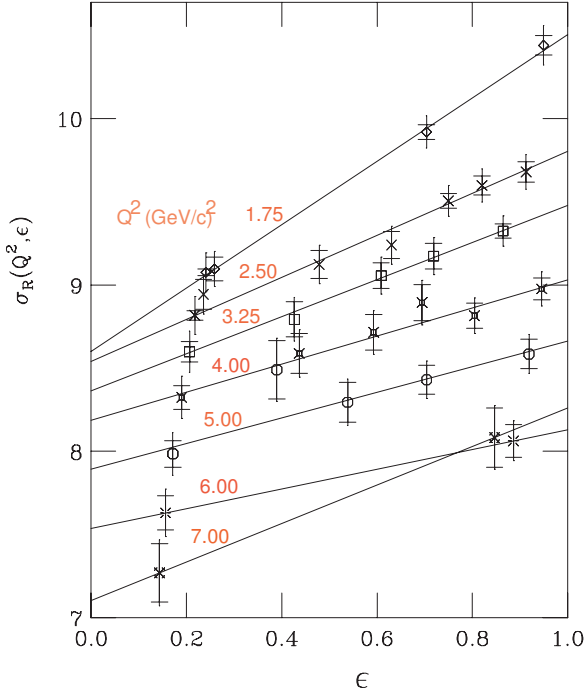}
    \end{minipage}\hfill
    \begin{minipage}[c]{0.48\textwidth}
      \caption{Reduced cross section plotted against $\epsilon$ for a range of fixed momentum transfers. The data is fit with a straight line with a slope $G_E^2$ and an intercept $\tau G^2_M$. The linear dependence of the reduced cross section  on $\epsilon$ assumes that any two-photon exchange effects are small. The data is from Ref.~\cite{Andivahis:1994rq}.}\label{fig:ne11}
    \end{minipage}
  \end{center}
\end{figure}

The proton form factors have been, until recently, only separated  through the Rosenbluth technique, which can be understood by re-writing Eq.~\ref{eq:f1f2} using the Sachs form factors,
\begin{equation}
\frac{d \sigma}{d \Omega} = \sigma_{\rm{NS}}
\left[\frac{G_E^2 + \tau G_M^2}{1+ \tau} +  2\tau G_M^2 \tan^2(\theta/2)\right],\end{equation} and rearranging, with $\epsilon^{-1} = 1 + 2(1+\tau)\tan^2(\theta/2)$ and $\sigma_{\rm{NS}}= \sigma _{\rm{Mott}}E'/E_{0}$, to give:

\begin{equation} \sigma_R \equiv \frac{d \sigma}{d \Omega} \frac{\epsilon(1+\tau)}{\sigma_{\rm{NS}}} = \tau G^2_M(Q^2)+\epsilon G_E^2(Q^2).
\end{equation} By making measurements at a fixed $Q^2$ and variable $\epsilon(\theta, E_0)$,
the reduced cross section $\sigma_{\rm{R}}$ can be fit with a straight line with slope $G_E^2$ and intercept $\tau G^2_M$.  Figure~\ref{fig:ne11} gives an example of the reduced cross section plotted in this way.
It should be noted that the Rosenbluth formula holds only for single photon exchange and it has been assumed (until recently) that any two-photon contribution is small.

\begin{figure}[!hb]
\centering   \includegraphics[width=3.1in]{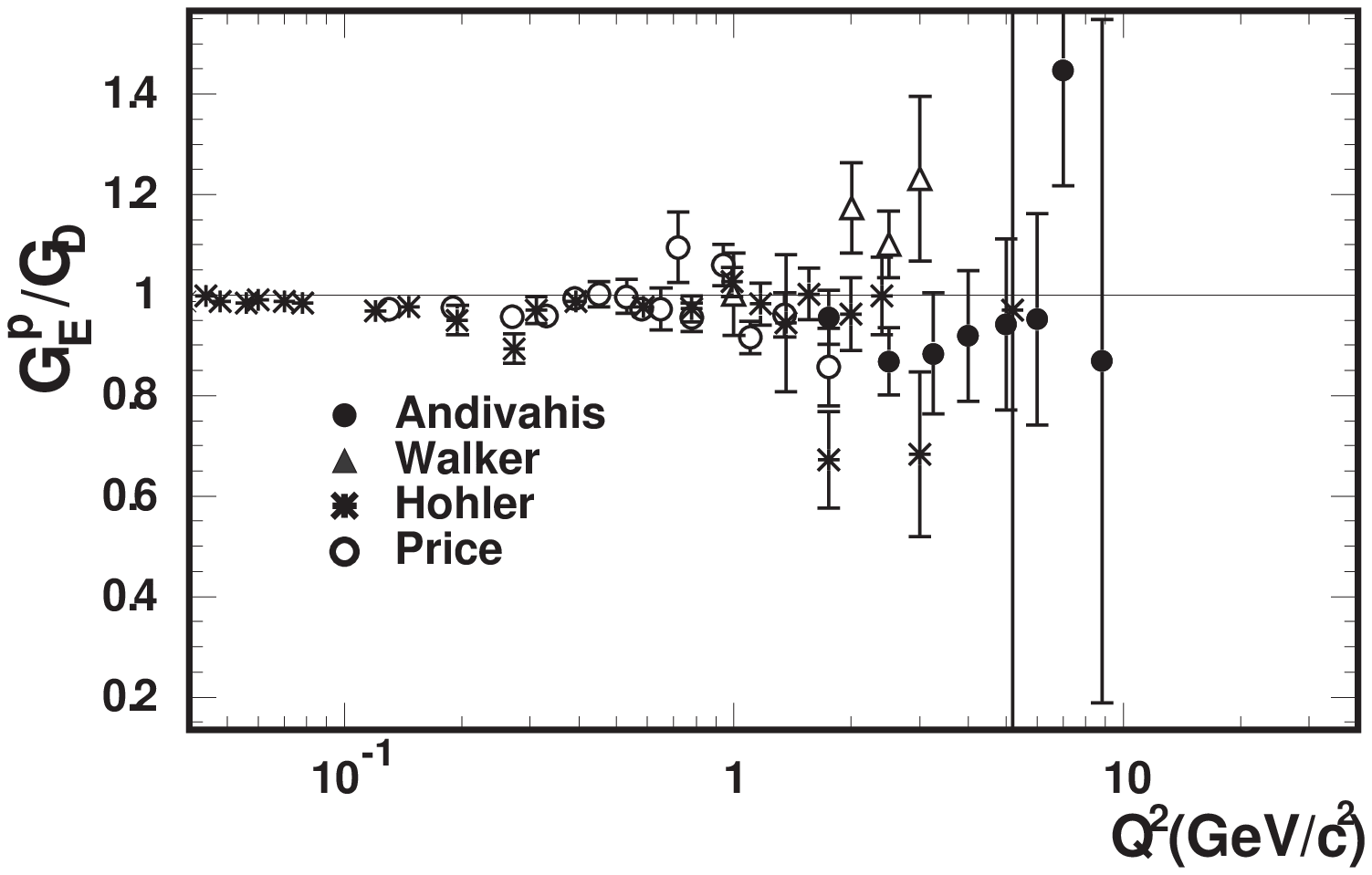}\hspace{0.05in}\includegraphics[width=3.1in]{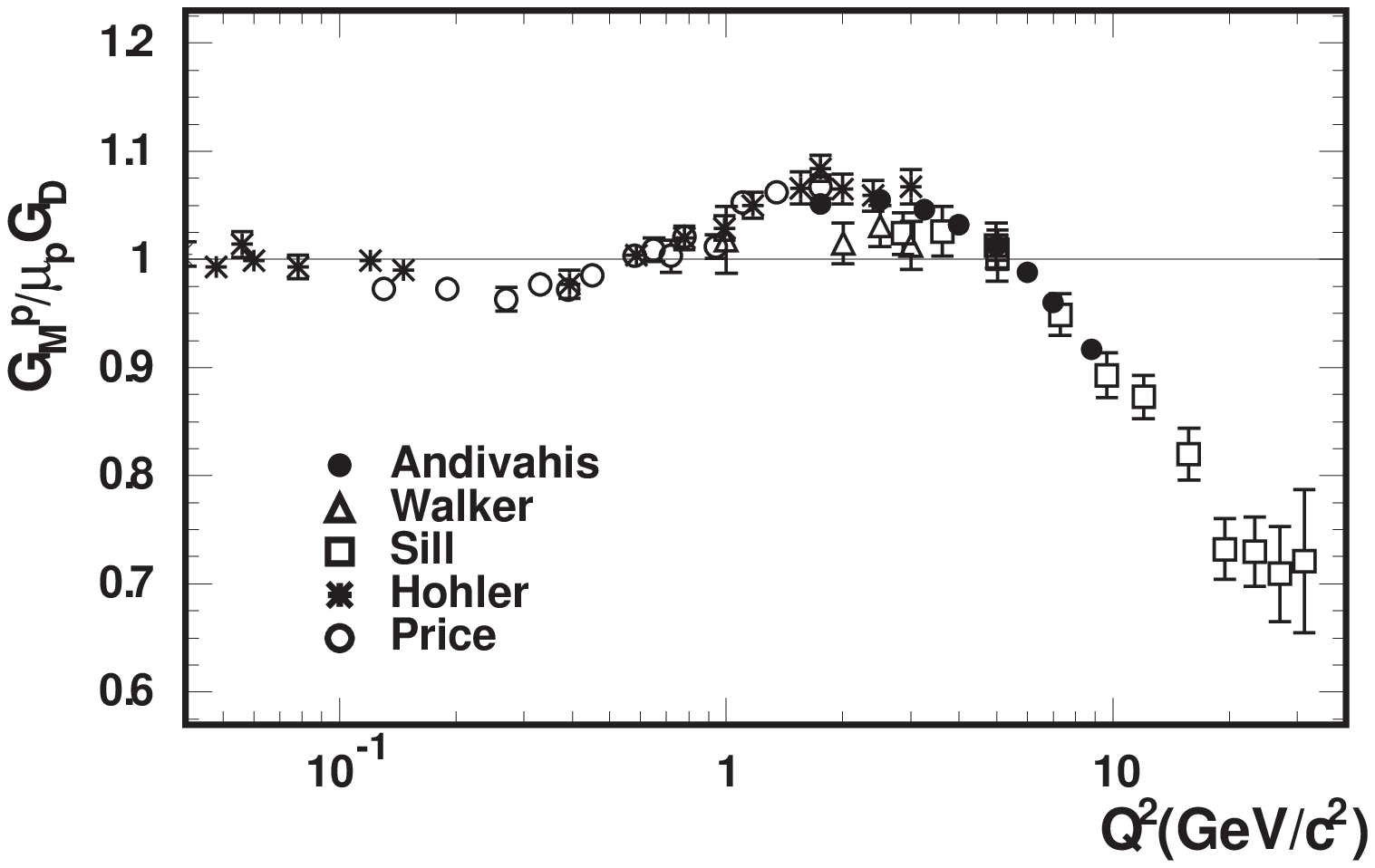}\\
  \caption{$G_E^p/G_D$ and $G_M^p/\mu_p/G_D$ versus $Q^2$ from the Rosenbluth method. The scaling law  and the dipole law hold to a good approximation ($\approx 10\%$) for both form factors out to $Q^2= 8$ (GeV/c)$^2$. }\label{fig:gepgmp}
\end{figure}

The Rosenbluth method is problematic--it requires the  measurement of absolute cross sections and  at  large $Q^2$ the cross section is insensitive to  $G_E$ and the error propagation suffers, $\delta G_E\propto \delta(\sigma_R(\epsilon_1) -\sigma_R(\epsilon_2))(\Delta\epsilon)^{-1}  (\tau G_M^2/G_E^2).$
Fig.~\ref{fig:gepgmp} presents the Rosenbluth data set (See Ref.~\cite{bosted95} for a compilation and references) for the  proton form factors. We see that the measurement of $G_M^p$ has been successfully made out to  30 (GeV/c)$^2$, while  the measurement of  $G_E^p$ begins to endure large errors at much smaller $Q^2$.
New measurements of the proton form factors exploiting a double polarization observable have upset the form factor scaling seen here.

\subsection{Neutron Form Factors}
\begin{figure}[!t]
\centerline {
\includegraphics[width=0.45\textwidth]{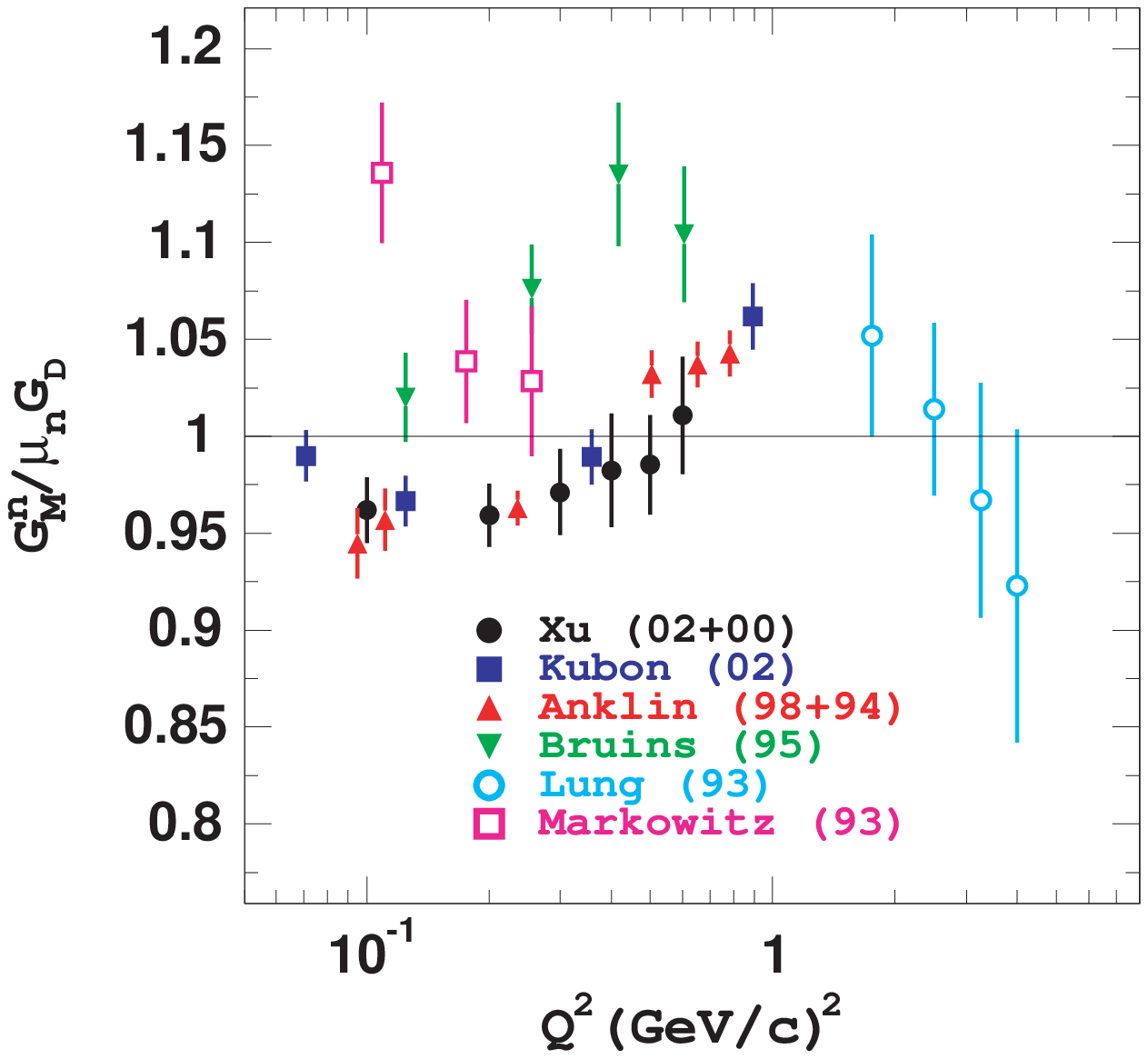}\includegraphics[angle=90,width=0.5\textwidth]{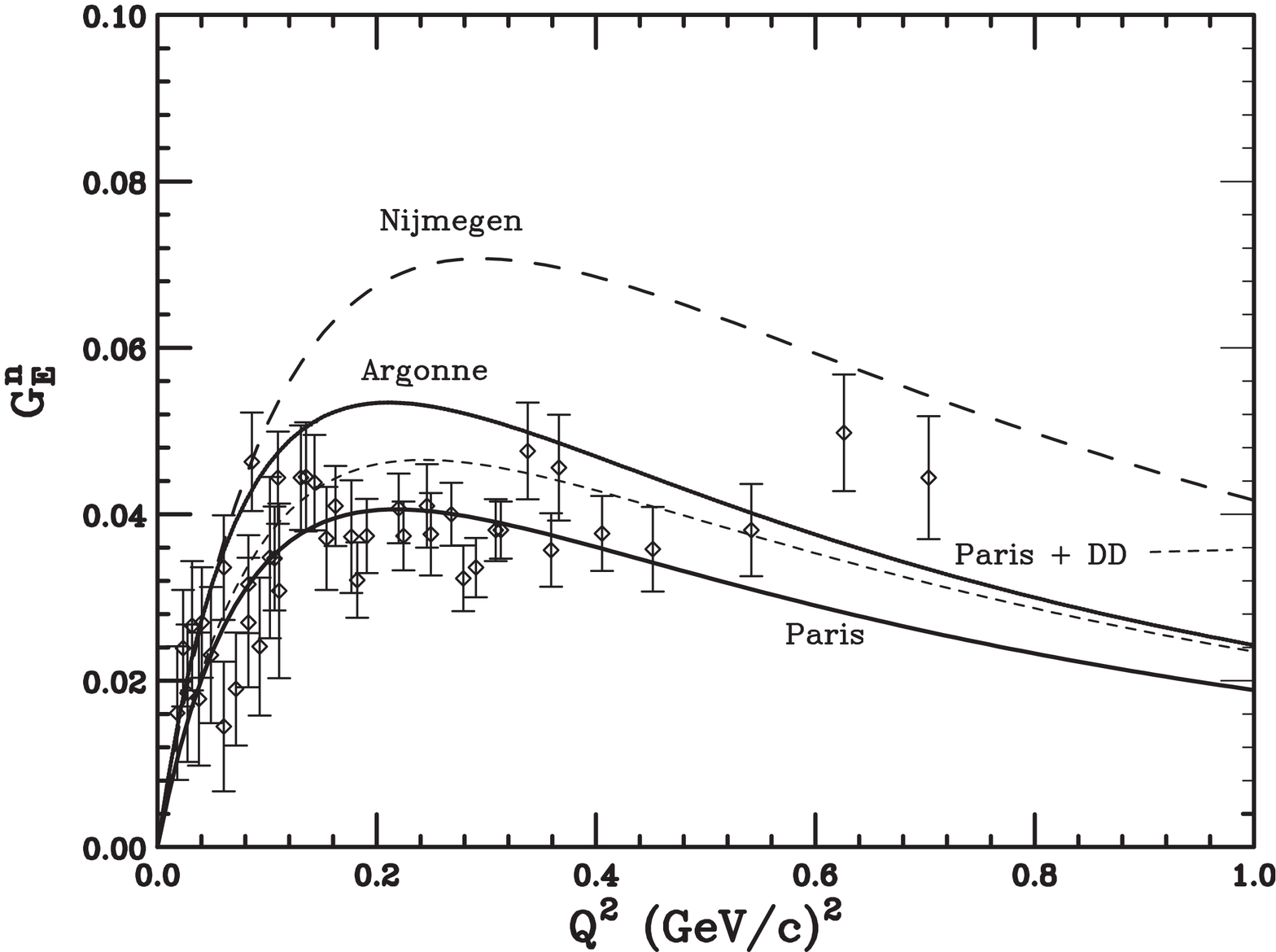}
} \caption[]{{Left: $G_M^n$  from unpolarized scattering~\cite{anklin94,bruins,anklin98,kubon02,lung93,mark93} and polarized scattering~\cite{xu00,Xu:2002xc}. Right: \gen from elastic e-D~\cite{platchov90}. The band defined by the lines represents the theoretical error associated with the extraction. The (small) dashed line is the familiar Galster parametrization\cite{Galster71}.\label{fig:gmngen}}}
\end{figure}

The lack  of a
free neutron target and the dominance of $G_M^n$ over $G_E^n$ (setting
aside recent progress) has left the data set on the
neutron form factors much less than desired.
The traditional
techniques (restricted to the use
of unpolarized beams and targets) used to  extract information about  $G_M^n$
and $G_E^n$ have been: elastic scattering from the deuteron (D):
 D$(e,e')$D;  inclusive quasielastic scattering:  D$(e,e')X$; 
scattering from deuteron with the coincident detection of the scattered electron
and recoiling neutron:  D$(e,e'n)p$;
 a ratio method which minimizes uncertainties in the deuteron
wave function and the role of FSI: $\frac{D(e,e' n)}{D(e,e'p)}$. The current status of the data on $G_M^n$ is shown in the LHS of Fig.~\ref{fig:gmngen}. The lack of data at large momentum transfers  will be soon remedied with data from an experiment~\cite{PR94-017,BrooksBaryons04} at Jefferson Lab which has used the ratio method to measure \gmn with small errors out to nearly 5 (GeV/c)$^2$ in the CLAS. This experiment took data on hydrogen and deuterium targets simultaneously. The neutron detector efficiency was determined in-situ via the $H(e,e'n\pi^+)$ reaction, and the CLAS's large acceptance allows the veto of any events with extra charged particles.

Until the early 1990's the extraction of \gen was done most successfully through
either small angle elastic $e$-D scattering~\cite{Galster71,platchov90} or
by quasielastic $e$-D scattering~\cite{lung93}.
In  the  Impulse Approximation (IA) the elastic electron-deuteron cross section is the sum of proton
and neutron responses with deuteron wave function weighting and in the small $
\theta _{e}$ approximation can be written,
\begin{equation}\frac{d \sigma}{d \Omega}\simeq \sigma_{\textrm{Mott}}(G_E^p + G_E^n)^2
\int_0^\infty\left[u(r)^2 + w(r)^2\right]j_0(\frac{qr}{2})dr. 
\end{equation}
The coherent nature of elastic scattering
gives rise to an interference term between the neutron and proton response which
allows the smaller \gen contribution to be extracted. Still,  the large proton
contribution must be removed. Experiments have been able to achieve small
statistical errors but remain very sensitive to deuteron wave function model
leaving a significant residual dependence on the NN potential.
The most precise data on \gen from elastic $e$-D scattering  are shown on the right hand side of Fig.~\ref{fig:gmngen} from an experiment at Saclay, published in 1990~\cite{platchov90}. The curves (parametrizations of the data based on different NN potentials used in the extraction) form a band which is a measure of the
theoretical uncertainty ($\approx$ 50\%) that cannot be avoided.

\section{SPIN DEPENDENT MEASUREMENTS}\label{sec:spindep}

It has been known for many years that  the nucleon electromagnetic form factors
could be measured through   spin-dependent elastic scattering from the nucleon, accomplished either through a measurement of the
scattering asymmetry of polarized electrons from a polarized nucleon target~\cite{dombey,Do89}
 or equivalently by measuring the polarization transferred to the nucleon~\cite{akhiezer,acg}.
In the scattering of polarized electrons from a polarized target, an asymmetry appears in the elastic scattering cross section when the beam helicity is  reversed. In contrast, in scattering a polarized electron from an unpolarized target, the transferred polarization to the nucleon produces an azimuthal asymmetry in the secondary scattering of the nucleon (in a polarimeter) due to its dependence on its polarization. In both cases, the perpendicular asymmetry
 is sensitive to the product $G_EG_M$.  Only in the last decade have experiments exploiting these spin degrees of freedom become possible.



Extraction of the neutron form factors (necessarily from a nuclear target) using polarization observables is complicated by the need to account for Fermi motion, MEC, and FSI, complications that are absent when scattering from a proton target. Fortunately it has been found for the deuteron that  in kinematics that emphasize quasi-free neutron knockout both the  transfer polarization $P_t$~\cite{arenhoevel87} and the beam-target asymmetry $A_V^{eD}$~\cite{arenhoevel88} are especially
sensitive to \gen  and  relatively  insensitive to the NN
potential describing the ground state of the deuteron and other reaction details. Calculations~\cite{golak} of the beam-target asymmetry from a polarized $^3$He target (which can be approximated as a polarized neutron) showed modest model dependence.

\subsection{Recoil polarization}
In elastic scattering of polarized electrons from a nucleon, the nucleon obtains (is transferred) a polarization whose components, $P_l$ (along the direction of the nucleon momentum)  and $P_t$ (perpendicular to the nucleon momentum) are proportional to $G_M^2$ and $G_EG_M$ respectively. The recoil polarization technique has allowed precision measurements of  $G_E^p$
to nearly 6 (GeV/c)$^2$~\cite{jones,gayou,gayou2} and of $G_E^n$ out to $Q^2 = 1.5 \rm{(GeV/c)}^2$~\cite{eden,He99,Os99,Madey:2003av}. Polarimeters are sensitive only to the perpendicular polarization components so precession of the nucleon spin before the polarimeter in the magnetic field of the spectrometer (for the proton) or a dipole (inserted in the path of neutron)  allows a measurement of the ratio $P_t/P_l$ and the form factor ratio: $\frac{G_E}{G_M} = - \frac{P_t}{P_l}\frac{(E_0+E')}{2M_N}\tan(\theta/2).$

\begin{figure}[t]
\includegraphics[width=0.5\textwidth]{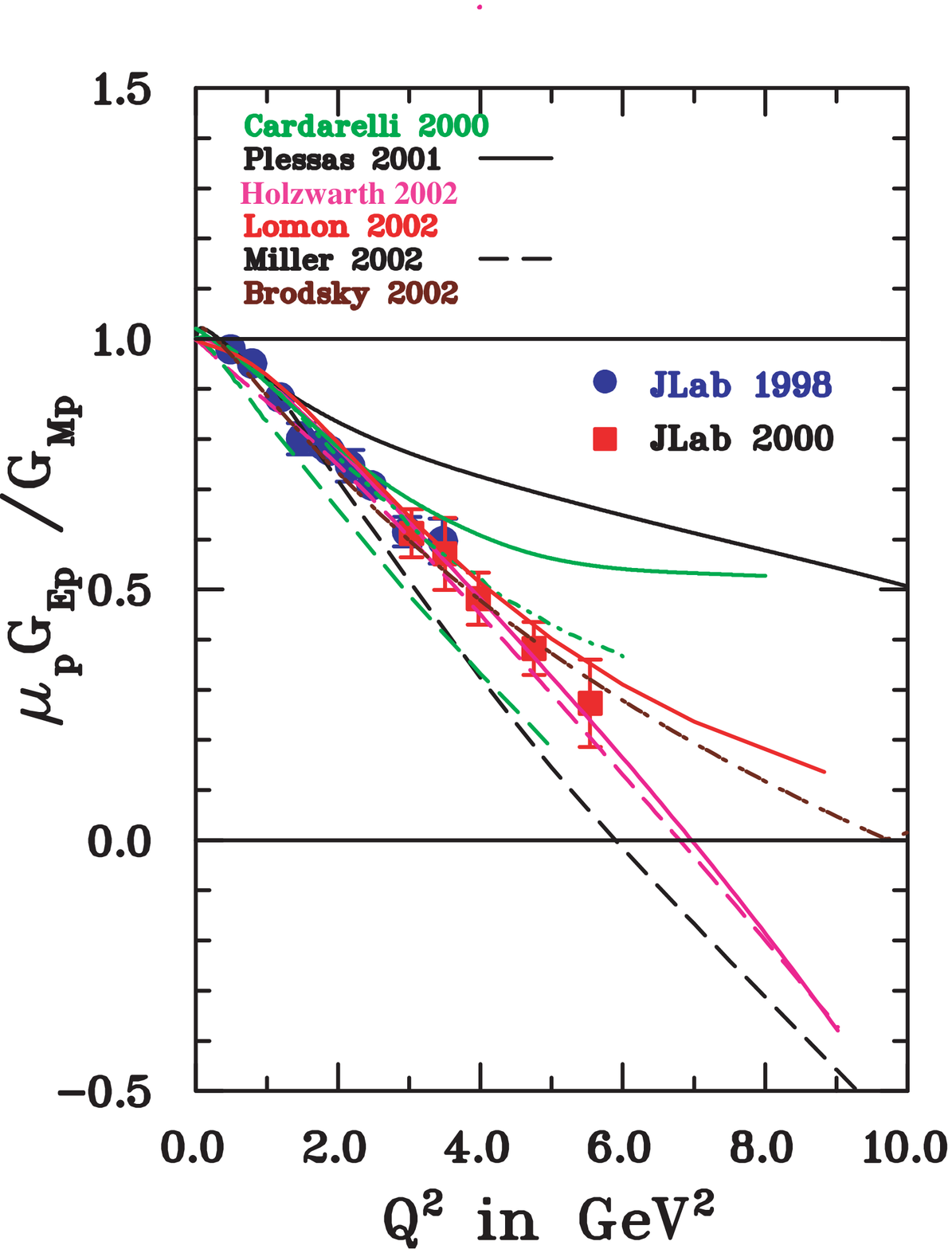}\includegraphics[width=0.5\textwidth]{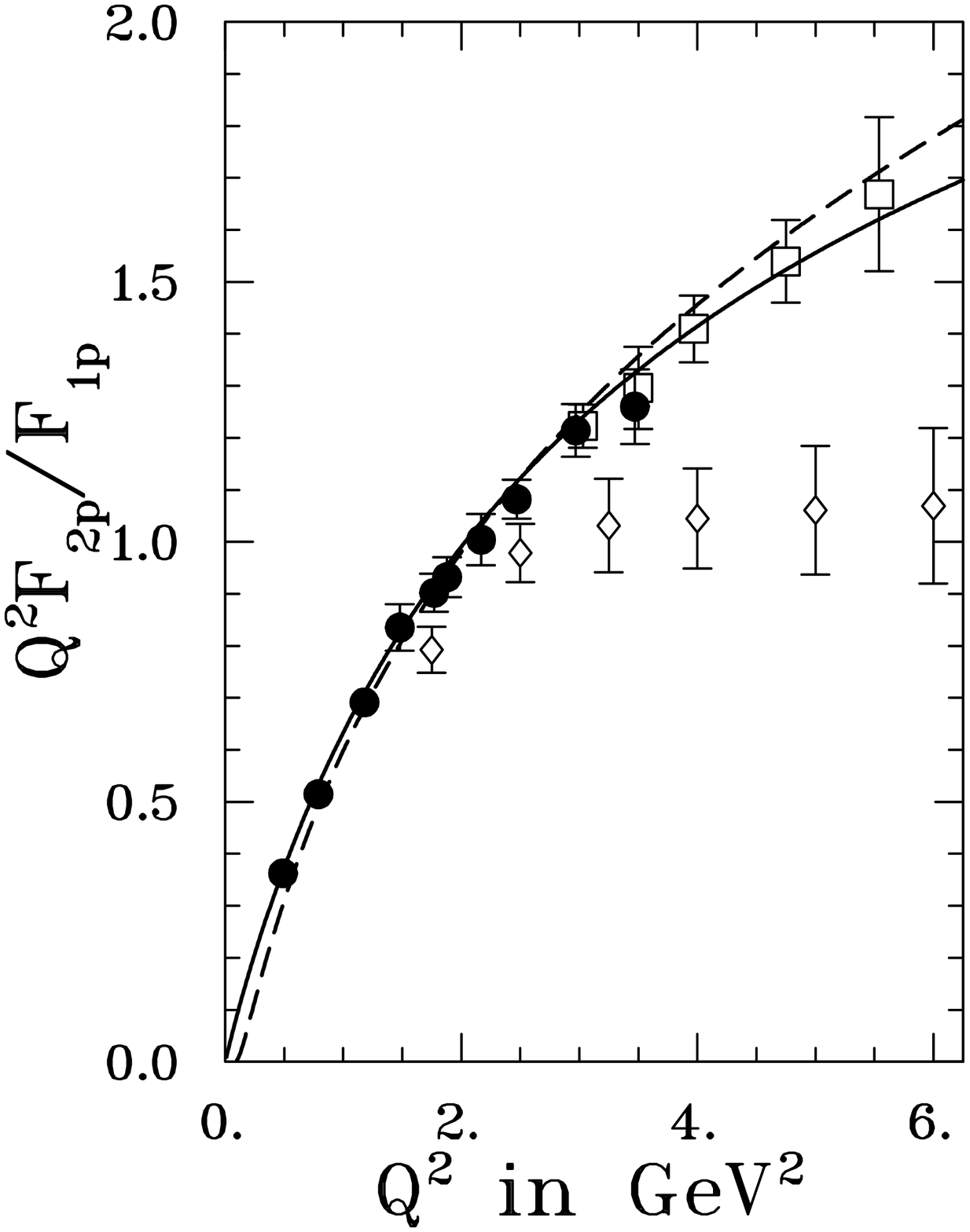}
\vspace{-0.35in}
\caption{Left: Comparison of theoretical model calculations with the data
from Ref.~\cite{jones} (solid circles) and from~\cite{gayou2} (solid squares). The curves are,
black thin solid~\cite{pfsa}, green dot-dashed and dashed~\cite{Pace:1999as}, black dashed~\cite{Frank1996}, red
solid \cite{Lomon:2002jx}, brown dashed~\cite{Brodsky:2002st} and magenta dashed and solid~\cite{holzwarth}.   Right: $Q^2F_2^p/F_1^p$ versus $Q^2$. The curves shown are from Brodsky \cite{Brodsky:2002st}(solid line), and from Belitsky et al.
\cite{Belitsky:2002kj} (dashed line). The data from Ref.~\cite{jones} are shown as solid circles, from
Ref.~\cite{gayou2} as empty squares and from Ref.~\cite{Andivahis:1994rq} as empty diamonds. For a discussion of the curves see Ref.~\cite{punjabi-03-conf} (from which these figures were taken).}\label{fig:jlabgep}
\end{figure}

The results from Jefferson Lab for the proton are shown in Fig.~\ref{fig:jlabgep} and it can immediately be seen that the  ratio of $\mu_pG_E^p/G_M^p$ does not follow the scaling law obtained from Rosenbluth separation, rather showing a steep decline with increasing $Q^2$ and suggesting that the distribution of magnetization and charge densities in the proton are dissimilar.  Shown with  the data are a collection of calculations including several relativistic Constituent Quark  Models (rQCM), a VMD-pQCD model and a chiral soliton model. Also shown is a pQCD calculation~\cite{Brodsky:2002st}. The right hand side of Fig.~\ref{fig:jlabgep} shows the same data, now in terms of $Q^2F_2/F_1$. The data give no indication of scaling at high momentum transfer, in contradiction to the early pQCD prediction~\cite{brodsky75-81}. The recent efforts~\cite{Brodsky:2002st}, still within pQCD and including higher twist contributions, have been able to reproduce this behavior (solid line). Other pQCD calculations  which consider quark angular orbital momentum are also successful, for example Refs.~\cite{Belitsky:2002kj,Ralston:2002ep}.

Recoil polarization has been used at both Jefferson Lab and Mainz to extract $G_E^n/G_M^n$ when scattering polarized electrons from an unpolarized deuteron target in quasi-elastic kinematics. At both labs a dipole magnet was used to precess the neutron spin thereby allowing a measurement of both polarization components. The results from \cite{Madey:2003av} are especially precise, extending our knowledge of \gen out to 1.5 GeV/c$^2$. See Fig.~\ref{fig:gendata}.

\subsection{Beam-target asymmetry}
Polarized targets have been used to extract \gen~\cite{Me94,Pa99,Be99,bermuth03,Glazier:2004ny,golak01,Zhu01,Warren:2003ma} and \gmn~\cite{gao94,xu00,Xu:2002xc}. The beam-target asymmetry can be written schematically ($a$, $b$, $c$, and $d$ are known kinematic factors) as    
\begin{equation}
A=\frac{ a  \cos\Theta^{\star} \left(G_{M}\right)^2  + 
          b  \sin\Theta^{\star} \cos\Phi^{\star} G_EG_M}
        {c \left(G_{M}\right)^2 + d \left(G_{E}\right)^2}
\end{equation} where $\Theta^{\star}$ and $\Phi^{\star}$ fix the target polarization axis.
 With the target polarization axis in the scattering plane and perpendicular to $\vec{q}$, ($\Theta^{\star}, \Phi^{\star}=90^{\circ},0^{\circ}$) the asymmetry $A_{TL}$ is proportional to $G_EG_M$. With the polarization axis in the scattering plane and parallel to $\vec{q}$ ($\Theta^{\star}, \Phi^{\star}=0^{\circ},0^{\circ}$), measuring the asymmetry $A_T$ allows $G_M$ to be determined. See the LHS of Fig.~\ref{fig:gmngen}.

\begin{figure}
\center\includegraphics[width=0.6\textwidth]{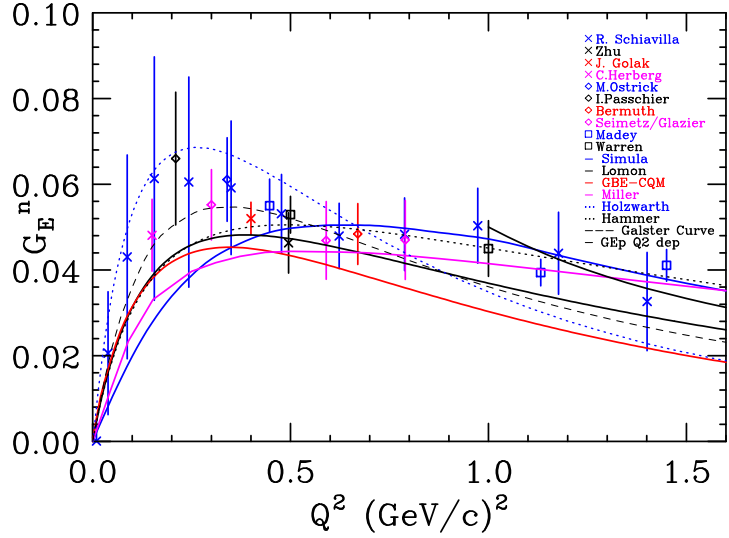}\vspace{-0.15in}
\caption[]{{Comparison of selected theoretical model calculations with the  data on \gen from polarized experiments.  Starting at the top of the legend the data are from~\cite{sick-sch01,Zhu01,golak01,He99,Os99,Pa99,bermuth03,Glazier:2004ny,Madey:2003av,Warren:2003ma}. Apparently, as can be seen by the  solid line starting at $Q^2$ = 1 (GeV/c)$^2$, the neutron, at large momentum transfers,  has the same  $Q^2$-dependence as the proton.}\label{fig:gendata}}
\end{figure}

 \gen has been extracted from beam-target asymmetry measurements using polarized $^3$He targets at Mainz and polarized ND$_3$ targets at Jefferson Lab,  and a polarized atomic beam target at NIKHEF. Data for \gen from both kinds of double polarization experiments are shown in Fig.~\ref{fig:gendata} along with some relevant calculations. The models, starting with the first in the  legend, include a rCQM ~\cite{Cardarelli2000}, a hybrid VMD-pQCD model~\cite{Lomon:2002jx}, a relativistic CQM calculation~\cite{pfsa}, a light-front cloudy bag model~\cite{miller}, a soliton model~\cite{holzwarth}, and a dispersion theory calculation~\cite{Hammer:2003ai}. While most of these calculations describe the $Q^2$-dependence, several badly fail to reproduce the slope at $Q^2=0, \frac{dG_E^n(0)}{dQ^2} = -\frac{1}{6} \left \langle r_{E}^2\right \rangle$. The neutron charge radius, $r_E$, has been determined through neutron electron scattering~\cite{Kopecki95}.

\section{THE $\mathbf {G_E^p/G_M^p\:}$ DISCREPANCY AND TWO-PHOTON EFFECTS}

The recoil polarization measurements of the form factor ratio $\mu_pG_E^p/G_M^p$  contradict the Rosenbluth measurements and it has been suggested  that the earlier experiments might have underestimated systematic errors or suffer from normalization problems. Recently the Rosenbluth measurements have been reexamined~\cite{Arrington:2003df}. This global reanalysis could find no systematic or normalization problems that could account for the discrepancy and concluded that a 5-6\% linear $\epsilon$-dependence correction (of origin yet unknown) to the cross section measurements is required to  explain the difference.  Several investigators~\cite{Blunden:2003sp,Rekalo:2003xa,Guichon:2003qm,Chen:2004tw} have explored the possibility of two-photon exchange corrections (which would be less important in the direct ratio measurement of recoil polarization) to explain the discrepancy. While only incomplete calculations exist, the results of Ref.~\cite{Blunden:2003sp,Chen:2004tw} account for part of the difference.

 The most recent work by Chen {\it{et al.}}~\cite{Chen:2004tw} employed a different approach than that of Ref.~\cite{Blunden:2003sp} in that they describe the process in terms of hard scattering from a quark and use GPD's to describe the quark emission and absorption. They argue that when taking the recoil polarization form factors as input, the addition of the two-photon corrections reproduces the Rosenbluth data. However, Arrington~\cite{Arrington:2004ae} has shown that when the corrections of  Chen {\it{et al.}} are applied to the new Jefferson Lab Rosenbluth data,  which have small errors, (see below and Figure~\ref{fig:super}) only one-half of the discrepancy is explained.

The Rosenbluth-polarization transfer discrepancy has been recently confirmed by a ``Super" Rosenbluth measurement \cite{E01-001} at Jefferson Lab that was designed to minimize the systematic errors that handicap Rosenbluth measurements.  This was achieved  by  detecting the proton rather than the electron in elastic kinematics. In doing so  many of the extreme rate variations and cross section sensitivities that are normally encountered  were avoided. The results \cite{Qattan:2004ht} show that the discrepancy still exists. See Figure~{\ref{fig:super}.

\begin{figure}[!ht]\includegraphics[width=0.5\textwidth]{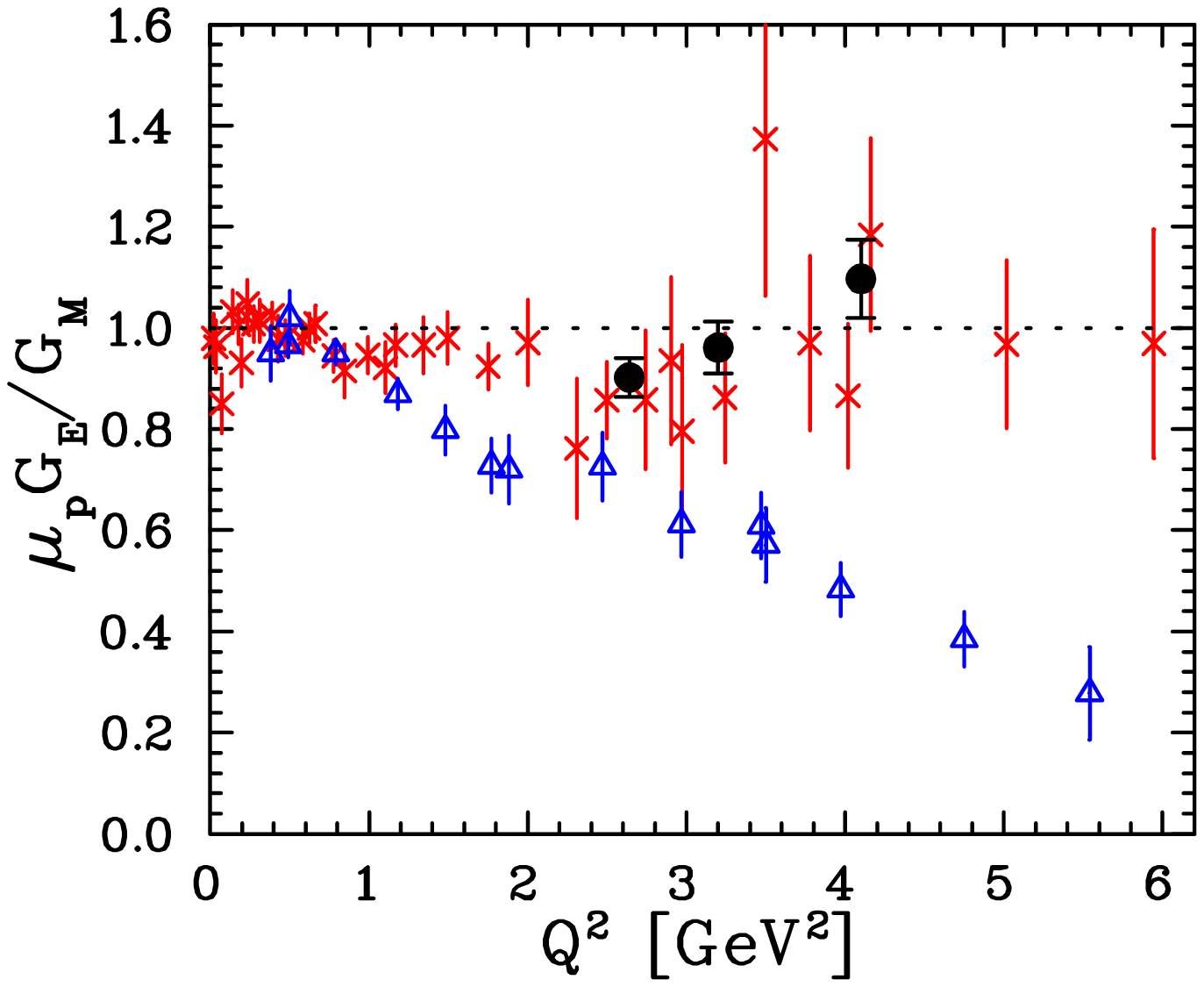}\includegraphics[width=.5\textwidth]{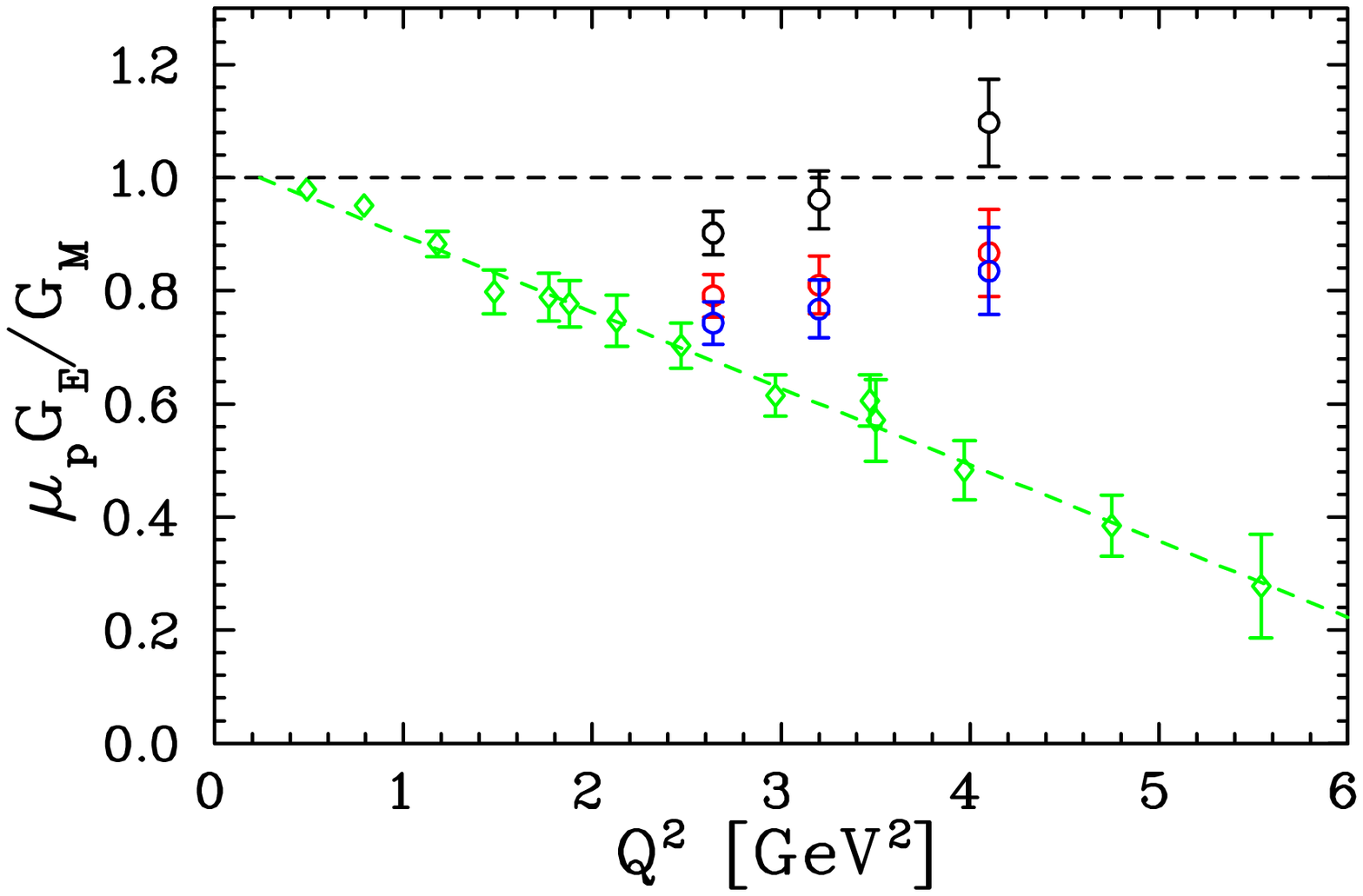}
\caption{ Left: Proton form factor ratio where the (blue) triangles are from recoil polarization~\cite{jones,gayou2}, (red) crosses from the reanalysis of the world's Rosenbluth data set~\cite{Arrington:2003df} and the filled circles from the recent super-Rosenbluth experiment~\cite{Qattan:2004ht}. Right: Corrections to the three data points from the super-Rosenbluth data set move those data towards,({\it i.e} reduces the value of the form factor ratio) the recoil polarization data set shown with fit (dotted line). Starting with the largest value at each $Q^2$ is the measured ratio, the ratio with the two-photon corrections of Ref.~\cite{Chen:2004tw} applied and the ratio  with  the two-photon and with Coulomb corrections~\cite{Arrington:2004is} applied~\cite{Arrington:2004ae}.}\label{fig:super}
\end{figure}

Direct experimental tests for the existence of two-photon exchange include  measurements of the ratio $\frac{\sigma(e^+p)}{\sigma(e^-p)}$, where the real part of the two-photon exchange amplitude leads to an enhancement, and in  Rosenbluth data where it can lead to non-linearities in $\epsilon$. There is no experimental evidence of non-linearities in the Rosenbluth data and the $e^+/e^-$ ratio data~\cite{Mar:1968qd} are of only modest precision, making it difficult to absolutely confirm the presence of two-photon effects in these processes.

It is the imaginary part of the two-photon amplitude that can lead to single spin asymmetries but again the  existing data~\cite{Kirkman:1970er,Powell:1970qt} are of insufficient precision to allow one to make a statement.   There is, however, one observable that has provided unambiguous evidence for a two-photon effect in $ep$ elastic scattering. Groups in both the US~\cite{Wells:2000rx} and Europe~\cite{Maas:2004pd}
have measured the transverse polarized beam asymmetry. These measurements are significant but have limited utility in solving the $G^p_E$ discrepancy. The reader interested in more detail about the existence of two-photon effects and their role on the form factor measurements should refer to Ref.~\cite{Arrington:2004ae}.

Fortunately  experiments are planned at Jefferson Lab to look for non-linearities in the Rosenbluth data, for the presence of induced recoil polarization and for an enhancement in the $e^+/e^-$ ratio. We can expect that a concentrated effort in both experiment and theory will reveal the full extent of two-photon effects in the not too distant future.
   
\section{OUTLOOK}

The capabilities of high duty factor accelerators, polarized beams and targets, and polarimeters have produced precision data out to large momentum transfer on the proton and neutron form factors. These new data, accumulated over the last 10 years, now challenge our previously held view of the structure of the proton  and neutron and provide a rigorous test for any QCD-inspired model of nucleon structure.


The conclusion from a review article \cite{RRWilson64} published 40 years ago is still appropriate today: {\it Although the major landmarks of this field of study are now clear, we are left with the feeling that much is yet to be learned about the nucleon by refining and extending both measurement and theory.}

\end{document}